\documentclass[iop,apjl,tighten,twocolappendix]{emulateapj}

\usepackage{apjfonts}

\usepackage{graphicx}
\usepackage[breaklinks,colorlinks=true,linkcolor=blue,citecolor=blue,urlcolor=blue]{hyperref}
\usepackage{bm}
\usepackage{color}
\usepackage{enumitem}
\usepackage{amsmath}
\usepackage[perpage]{footmisc}
\usepackage{multirow}
\usepackage{booktabs}
\usepackage{array}
\usepackage{dcolumn}
\usepackage{tablefootnote}
\usepackage[table]{xcolor}
\usepackage{colortbl}

\usepackage{CJK}

\usepackage{etoolbox}
\makeatletter
\patchcmd{\NAT@citex}
  {\@citea\NAT@hyper@{%
     \NAT@nmfmt{\NAT@nm}%
     \hyper@natlinkbreak{\NAT@aysep\NAT@spacechar}{\@citeb\@extra@b@citeb}%
     \NAT@date}}
  {\@citea\NAT@nmfmt{\NAT@nm}%
   \NAT@aysep\NAT@spacechar\NAT@hyper@{\NAT@date}}{}{}
\patchcmd{\NAT@citex}
  {\@citea\NAT@hyper@{%
     \NAT@nmfmt{\NAT@nm}%
     \hyper@natlinkbreak{\NAT@spacechar\NAT@@open\if*#1*\else#1\NAT@spacechar\fi}%
       {\@citeb\@extra@b@citeb}%
     \NAT@date}}
  {\@citea\NAT@nmfmt{\NAT@nm}%
   \NAT@spacechar\NAT@@open\if*#1*\else#1\NAT@spacechar\fi\NAT@hyper@{\NAT@date}}
  {}{}
\makeatother

\shorttitle{AR 12192}
\shortauthors{Sun et al.}

\slugcomment{Accepted for publication in the Astrophysical Journal Letters} 

\begin{document}

\begin{CJK}{UTF8}{}

\title{
Why Is the Great Solar Active Region 12192 Flare-Rich But CME-Poor?
}

\author{
\begin{CJK}{UTF8}{gbsn} % Use default fonts from CJK (see below)
Xudong Sun (孙旭东)\altaffilmark{1}, Monica G. Bobra\altaffilmark{1}, J. Todd Hoeksema\altaffilmark{1}, Yang Liu (刘扬)\altaffilmark{1}, Yan Li\altaffilmark{2},\\[2pt] Chenglong Shen (申成龙)\altaffilmark{3}, Sebastien Couvidat\altaffilmark{1}, Aimee A. Norton\altaffilmark{1}, and George H. Fisher\altaffilmark{2}
\end{CJK}
}

\affil{
$^1$W. W. Hansen Experimental Physics Laboratory, Stanford University, Stanford, CA 94305-4085, USA; \href{mailto:xudong@sun.stanford.edu}{xudong@sun.stanford.edu}\\
$^2$Space Sciences Laboratory, University of California, Berkeley, CA 94720-7450, USA\\
$^3$School of Earth and Space Sciences, University of Science and Technology of China, Hefei 230026, China
}

%%%%%%%%%%%%%%%%%%%

\begin{abstract}
Solar active region (AR) 12192 of October 2014 hosts the largest sunspot group in 24 years. It is the most prolific flaring site of Cycle 24 so far, but surprisingly produced no coronal mass ejection (CME) from the core region during its disk passage. Here, we study the magnetic conditions that prevented eruption and the consequences that ensued. We find AR 12192 to be ``big but mild''; its core region exhibits weaker non-potentiality, stronger overlying field, and smaller flare-related field changes compared to two other major flare-CME-productive ARs (11429 and 11158). These differences are present in the intensive-type indices (e.g., means) but generally not the extensive ones (e.g., totals). AR 12192's large amount of magnetic free energy does not translate into CME productivity. The unexpected behavior suggests that AR eruptiveness is limited by some \textit{relative} measure of magnetic non-potentiality over the restriction of background field, and that confined flares may leave weaker photospheric and coronal imprints compared to their eruptive counterparts.
\end{abstract}


%%%%%%%%%%%%%%%%%%%
%%%%%%%%%%%%%%%%%%%

\section{Introduction}
\label{sec:intro}

The great solar active region (AR) 12192 of October 2014 harbors the largest sunspot group in 24 years. It is so far the most intensely flaring region of Cycle 24, producing a total of six \textit{GOES} X-class flares and a multitude of smaller ones. Statistically, both flares and coronal mass ejections (CMEs) tend to occur in these intense events. A survey of 1996--2005 indicates that 75 of 90 X-class flares are associated with CMEs; for those above X3 the rate is 23 in 24 \citep{yashiro2006}. It is therefore surprising that no CME was detected from the core region of AR 12192 during its disk passage. Only one on-disk jet-like CME originated from the AR periphery; two others have been reported to erupt from over the east limb \citep{west2015}. The unexpected behavior quickly raised interest from the community (e.g., \textit{RHESSI} Science Nugget \href{http://sprg.ssl.berkeley.edu/~tohban/wiki/index.php/A_Record-Setting_CMEless_Flare}{\#239}; \citealt{thalmann2015}).

We dub all flares without a CME ``confined'': they either produce no eruption, or eruptions that fail to escape the Sun \citep[e.g.,][]{ji2003}. Confined flares are rare for the more energetic events; only a dozen or so confined X-class flares have been analyzed \citep{schmahl1990,feynman1994,gaizauskas1998,green2002,wangyuming2007,liuyang2008,cheng2011,chenhd2013,liur2014}. On the other hand, the lack of an associated CME is common among weaker ones, accounting for 40$\%$ of M-class flares \citep{andrews2003} and a majority for C-class and below. ARs with large eruptive flares often produce smaller, confined events too.

The magnetic cause of confined flares has been studied along with the eruption mechanism. Comparative case studies have largely probed two aspects. One focus is AR non-potentiality as the source for eruption, e.g., magnetic helicity \citep{nindos2004} and kink instability \citep{guo2010}. The other focus is the constraining effect of the background field, e.g., its decay with height \citep{liuyang2008,guo2010,cheng2011,nindos2012} and its strength \citep{wangyuming2007,liuyang2008}. Numerical experiments seem to suggest that both non-potentiality and the background field contribute: by fixing one and adjusting the other, confined events can transition to eruptions \citep{amari2003,torok2005}. Statistical studies have shown good correlation between CME occurrence and magnetic twist, electric current, and global free energy proxies \citep{falconer2002,falconer2006}. Recent work suggests an ``upper limit'' on free energy, where major flares and CMEs preferentially occur \citep{falconer2009,moore2012}.

The magnetic consequence of confined flares is little explored in comparison. A CME bodily removes twisted magnetic structure, resulting in decrease of magnetic helicity. The flare-related, stepwise change in the photospheric field \citep[e.g.,][]{sun2012,wangs2012} have been interpreted to be a record of the Lorentz force impulse that drives the ejecta \citep{fisher2012}. Without a CME, do confined flares yield smaller magnetic field changes compared to their eruptive counterparts?

Closely watched by multiple observatories, AR 12192 is a wonderful test case with all its peculiar behaviors. Here, we utilize photospheric vector magnetic data from the Helioseismic and Magnetic Imager \citep[HMI;][]{schou2012} aboard the \textit{Solar Dynamics Observatory} (\textit{SDO}) to probe the cause and consequence of the largest confined flare (X3.1) from AR 12192. To this end, we select two additional ARs (11429 and 11158) with major eruptive flares and compare their pre- and post-explosion magnetic conditions. Several distinctive features of AR 12192 immediately stand out. We discuss the implications of our observation.

%%%%%%%%%%%%%%%

\begin{figure*}[t!]
\centerline{\includegraphics{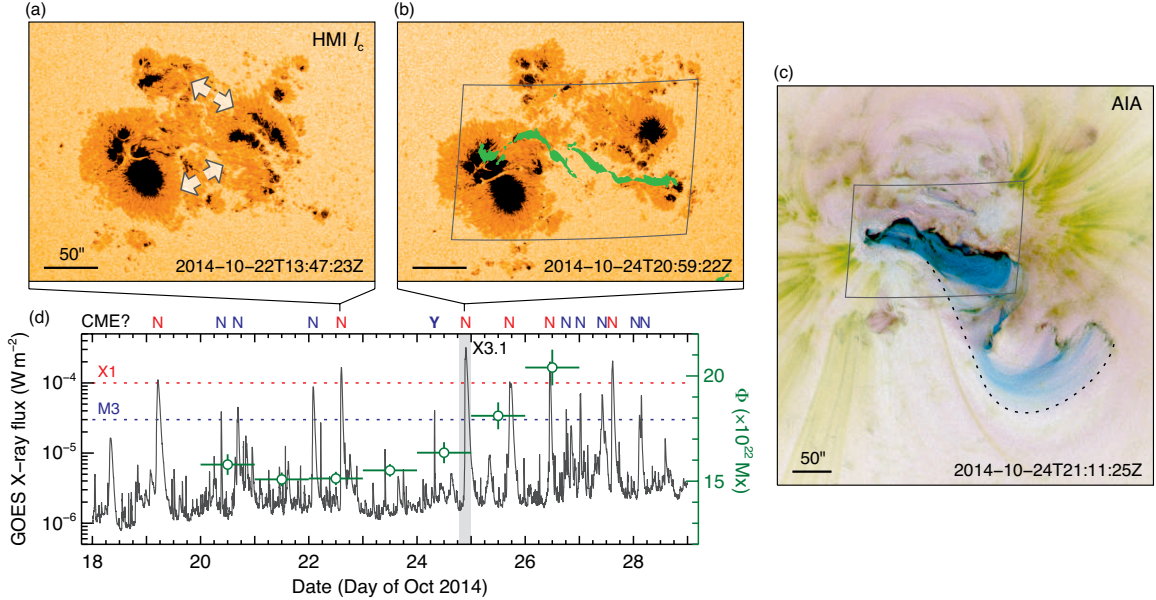}}
\caption{Overview of AR 12192. (a) (b) HMI continuum intensity before two X-class flares with scattered light removed. Arrows in (a) denote two locations where significant sunspot separation took place. AIA 1600 {\AA} ribbons for the X3.1 flare are overplotted in (b). (c) Negative composite AIA image during the X3.1 flare. Cyan, yellow, and magenta show 131, 171 and 335 {\AA} passbands, respectively. Dotted curve outlines the hot 131 {\AA} loops connecting the main flaring site to the southwest. Boxes in (b) (c) define the extent of Figures~\ref{f:compare}(a). (d) \textit{GOES} 1--8 {\AA} flux (black) and unsigned magnetic flux $\Phi$ (green symbols). The AR passed central meridian on October 23. CME production of each major flare is marked along the top. Horizontal error bars of $\Phi$ indicate 1-day window; vertical error bars indicate daily standard deviation.}
\label{f:context}
\end{figure*}

%%%%%%%%%%%%%%%%%%%

\renewcommand{\arraystretch}{1.25}
\newcommand*{\ma}[1]{\multicolumn{1}{c}{#1}}
%\definecolor{G}{gray}{0.85}

\begin{table*}[t!]
\begin{center}
\caption{Comparison of magnetic characteristics of three major active regions.\label{t:keys}}
\begin{tabular}{llp{22mm}D{,}{\pm}{-1}D{,}{\pm}{-1}D{,}{\pm}{-1}lp{6mm}}
\toprule
& & & \ma{AR 12192} & \ma{AR 11429} & \ma{AR 11158} & Unit & Type$^\ast$\\
\midrule
\parbox[t]{2mm}{\multirow{7}{*}{\rotatebox[origin=c]{90}{Flare and CME$^\dag$}}} & & Flare index & \ma{2335} & \ma{1295} & \ma{592} & \\
    & & Major flares & \ma{15} & \ma{7} & \ma{3} & \\ [2.4pt]
    \cmidrule{3-6}
    & & Event & \ma{SOL2014-10-24T21:41} & \ma{SOL2012-03-07T00:24} & \ma{SOL2011-02-15T01:56} \\ 
    & & Location & \ma{S21W21} & \ma{N18E31} & \ma{S20W10} & \\
    & & \textit{GOES} class & \ma{X3.1} & \ma{X5.4} & \ma{X2.2} & \\
    & & Duration & \ma{66} & \ma{38} & \ma{22} & min \\
    & & CME & \ma{No} & \ma{Halo} & \ma{Halo} & \\
\midrule
\parbox[t]{2mm}{\multirow{12}{*}{\rotatebox[origin=c]{90}{Photosphere$^\ddag$}}} & \parbox[t]{2mm}{\multirow{4}{*}{\rotatebox[origin=c]{90}{Overall}}} & Sunspot area & 4002,11 & 1490,2 & 861,4 & $\mu$Hem & E \\
 & & $\Phi$ & 16.12,0.08 & 4.88,0.04 & 2.73,0.04 & $10^{22}$ Mx & E \\ 
 & & $I$ & 25.98,0.00 & 8.00,0.00 & 6.31,0.00 & $10^{13}$ A  & E \\
 & & $\log R$ & 5.30,0.01 & 5.32,0.01 & 4.89,0.01 &  & E \\ [2.4pt]
 \cmidrule{3-6}
 & \parbox[t]{2mm}{\multirow{8}{*}{\rotatebox[origin=c]{90}{FPIL}}} & Mask area & 357,4 & 231,2 & 224,10 & $\mu$Hem & E \\
 & & $\Phi$ & 0.31,0.03 & 0.45,0.03 & 0.28,0.04 & $10^{22}$ Mx & E \\
 & & $I$ & 1.10,0.00 & 1.35,0.00 & 1.12,0.00 & $10^{13}$ A & E \\
 & & $B$ (rms) & 453,3 & 827,4 & 678,14 & G & I \\
 & & $J$ (rms) & 16.1,0.3 & 30.7,0.6 & 27.7,0.4 & mA m$^{-2}$ & I \\
 & & Shear & 50.2,0.3 & 57.9,0.3 & 60.1,0.7 & degree & I \\
 & & $|\alpha|$ & \textbf{0.05},\textbf{0.00} & \textbf{0.19},\textbf{0.00} & \textbf{0.25},\textbf{0.01} & Mm$^{-1}$ & I \\
 & & Current helicity & 0.72,0.04 & 6.74,0.06 & 5.28,0.20 & G$^2$ m$^{-1}$ & E \\
\midrule
\parbox[t]{2mm}{\multirow{6}{*}{\rotatebox[origin=c]{90}{Low corona$^\parallel$}}} & \parbox[t]{2mm}{\multirow{3}{*}{\rotatebox[origin=c]{90}{Overlying}}} & $B_h(42)$ & 220,8 & 61,7 & 42,0 & G & I \\
 & & $B_h(42)/B_h(2)$ & \textbf{0.35},\textbf{0.04} & \textbf{0.06},\textbf{0.00} & \textbf{0.05},\textbf{0.00} &  & I \\
 & & Critical height & 77,1 & 34,0 & 42,1 & Mm & I \\ [2.4pt]
 \cmidrule{3-6}
 & \parbox[t]{2mm}{\multirow{3}{*}{\rotatebox[origin=c]{90}{Energy}}} & $E_p$ & 152.8,0.2 & 20.9,0.1 & 8.8,0.0 & $10^{32}$ erg & E \\
 & & $E_f$ & 4.5,0.0 & 10.6,0.0 & 2.5,0.0 & $10^{32}$ erg & E \\
 & & $E_f/E_p$ & \textbf{0.03},\textbf{0.00} & \textbf{0.51},\textbf{0.02} & \textbf{0.28},\textbf{0.01} & & I \\
\midrule
\parbox[t]{2mm}{\multirow{5}{*}{\rotatebox[origin=c]{90}{Change$^\S$}}} & & $\Delta E_f$ & \ma{-0.90} & \ma{-1.58} & \ma{-0.26} & 10$^{32}$ erg & E \\
 & & $\Delta (\sum B_h \delta A)$ & \ma{+1} & \ma{+14} & \ma{+8} & 10$^{20}$ Mx & E \\
 & & $\Delta \left< B_h \right>$ & \ma{\textbf{+11}} & \ma{\textbf{+200}} & \ma{\textbf{+129}} & G & I \\
 & & $\Delta F_z$ & \ma{+0.2} & \ma{+11.1} & \ma{+4.7} & 10$^{22}$ dyne & E \\
 & & Topology change & \ma{Small} & \ma{Large} & \ma{Large} & & I \\
\bottomrule
\end{tabular} \\ [2mm]
%\parbox[c]{15.1cm}{
\parbox[c]{15.9cm}{
{\textbf{Notes.}}\\
{$^\ast$Indices are classified as extensive (E) or intensive (I); see Section~\ref{subsec:preflare}. Indices in bold are arbitrarily selected as examples for each category.}\\
{$^\dag$Flare index is defined as $\sum100M_X+\sum10M_M+\sum M_C$, where $M_X$ indicates the \textit{GOES} magnitude of each X-class flare, etc. Major flares include those above M3, between E70 and W70.}\\
{$^\ddag$Sunspot area is computed from HMI intensity, including both umbrae and penumbrae. $R$ measures the total unsigned flux within 15 Mm of high-gradient PIL \citep{schrijver2007}, here with $B_z$ instead of line-of-sight maps. Mean shear is the mean angle between the observed and the modeled PF on the photosphere; mean torsional parameter $\alpha$ is calculated as $\sum B_z J_z / \sum B_z^2$; current helicity is approximated by $|\sum B_z J_z|$ \citep{bobra2014}.}\\
{$^\parallel$The overlying field refers to $B_h$ directly above the FPIL in the PF model. $B_h(42)$ indicates mean $B_h$ at $42\pm1$ Mm, typical height of eruption onset \citep{liuyang2008}. $B_h(42)/B_h(2)$ is the mean ratio of $B_h$ at $42\pm1$ and $2\pm1$ Mm \citep[cf.][]{wangyuming2007}. The critical height is where the $B_h$ decay index $n$ reaches 1.5 so the torus instability may set in \citep{kliem2006}.}\\
{$^\S$The change of the surface integral $\sum B_h \delta A$ and the mean $\left< B_h \right>$ consider the FPIL region only, where $\delta A$ is the pixel area. The change of ``Lorentz force'' $F_z$ refers to the change of $\sum (B_h^2-B_z^2)\delta A/(8\pi)$ within FPIL \citep{fisher2012}. Topological change is assessed qualitatively based on $Q$ and coronal field connectivity (Figures~\ref{f:change}).}
}
\end{center}
\end{table*}

%%%%%%%%%%%%%%%%%%%

\section{Observation and Analysis}
\label{sec:obs}

AR 12192 was well developed when it rotated into view on 2014 October 17. Subsequent photospheric evolution mainly involves fast sunspot separation at two locations (Figure~\ref{f:context}(a)). Significant flux emergence commenced at the northern site when the AR was near central meridian. The sunspot area (HMI continuum intensity below 0.9 quiet Sun value) reached a maximum of $\sim$4300 $\mu$Hem. The unsigned flux $\Phi$ (sum of all pixels where field strength $B>200$ G) reached $\sim$2$\times$10$^{23}$ Mx on October 27 (Figure~\ref{f:context}(d)), an order of magnitude greater than the typical total net flux in the polar region during activity minimum \citep{sun2015}. The typical ratio of the total signed to unsigned flux is 0.08, suggesting a largely closed-field environment. All of the confined flares above M3 took place in the core region and showed double chromospheric flare ribbons (Figure~\ref{f:context}(b)). The overlying loops to the southwest appear to be directly involved in many events (Figure~\ref{f:context}(c)).

We aim to compare the magnetic condition of AR 12192 around SOL2014-10-24T21:41 (X3.1) with several other ARs of Cycle 24 that produced major eruptions. We search the \textit{GOES} flare list between 2010 and 2014 using the following criteria: their peak intensity is greater than X2; they occur between E40 and W40; they produce wide CMEs (halo or width greater than 60$^\circ$); and they show clear, extended double ribbons. The last criterion reduces the effect of complex magnetic topology; it rules out two candidates with fan-spine structure and circular-shaped ribbons \citep[e.g.,][]{sun2013}. Finally, the two regions selected are AR 11429 at SOL2012-03-07T00:24 (X5.4) and AR 11158 at SOL2011-02-15T01:56 (X2.2). The former is the second-most flare-productive AR of Cycle 24 \citep[e.g.,][]{wangr2014,liuying2014}. The latter produced the first X-class flare of Cycle 24 and is well studied \citep[e.g.,][]{sun2012}. All flares above M3 from these two ARs were associated with CMEs. Both ARs also produced multiple weaker, confined flares. The opposite-polarity sunspots in both ARs underwent strong shearing, in contrast to AR 12192.

HMI generates full disk photospheric vector magnetograms with 0.5$\arcsec$ plate scale at 12-min cadence. Each magnetogram is made from filtergrams taken within a $\sim$22.5 min averaging window, which has a uniform weight for the center and less contribution from the edges \citep{hoeksema2014}. ARs are automatically identified and extracted; de-projected maps are provided in cylindrical equal area coordinate \citep{bobra2014}. Here, we use five maps prior to the onset of each event to represent the pre-flare condition, and another five one hour after for the post-flare condition. Considering the wide averaging window, we make sure that the last pre-flare maps are centered at least 8 min before the flare onset. All contributing filtergrams are thus taken well before the hard X-ray peak to avoid possible artifacts from the intense flare emission \citep{qiu2003}. When possible, we consider only strong field regions where $B$ is greater than 200 G; results are presented as mean $\pm$ standard deviation of the five measurements.

We study the coronal field by extrapolating a nonlinear force-free field \citep[NLFFF;][]{wiegelmann2004,wiegelmann2006} and a potential field \citep[PF; e.g.,][]{sakurai1989} from each vector magnetogram \citep{sun2012}. Magnetic energy is calculated as $E=\sum B^2 \delta V / 8 \pi$, where $\delta V$ is the grid volume. The difference of the NLFFF energy $E_n$ and the PF energy $E_p$ indicates the free energy $E_f$ that is available to power the explosions. We note that $E_f$ should be taken as an order of magnitude estimate as the systematic uncertainty can be large \citep{sun2012}. The decay index, $n=-\partial \ln B_h / \partial \ln z$, characterizes the decrease rate of the horizontal field $B_h$ with height $z$ \citep[e.g.,][]{kliem2006}. We also compute the squashing factor $Q$ \citep{demoulin1996,pariat2012} to highlight the topological boundaries in the modeled field.

\begin{figure*}[t!]
\centerline{\includegraphics{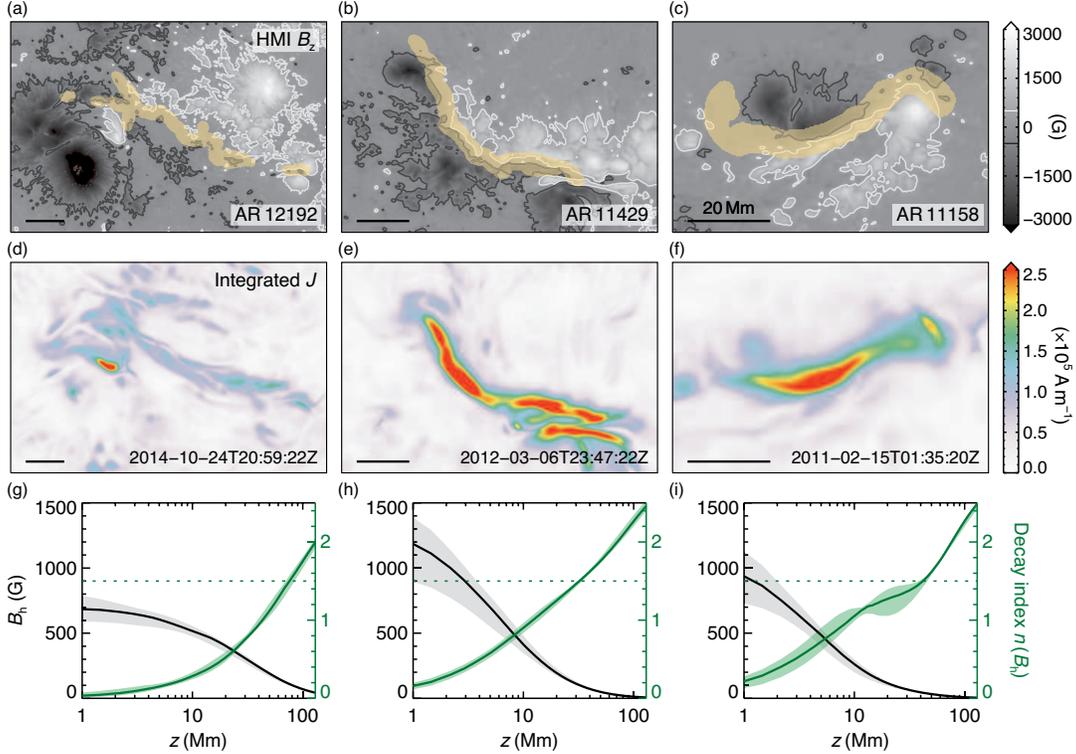}}
\caption{Comparison of pre-flare magnetic conditions of AR 12192, 11429, and 11158 prior to their respectively largest flare. (a)-(c) $B_z$ maps of the core region. They are 164, 120, and 76 Mm respectively in width. The yellow shaded regions denote our FPIL masks. (d)-(f) Maps of the vertically integrated $J$ over the lower 11 Mm in the NLFFF model. (g)-(i) Height profile of $B_h$ (black) and decay index $n$ (green) above the FPIL in the PF model. A total of 1--2$\times$10$^3$ profiles are evaluated for each AR; outliers are removed by using a K-mean algorithm. Lines show the median; shaded bands indicate $1\sigma$ spread. Horizontal dotted line indicates the critical value $n=1.5$. In (i), the kink \citep[cf.][]{nindos2012} and the larger spread of $n$ are due to the quadrupolar nature of AR 11158 (only the central bipole is shown): magnetic connectivity changes rapidly at 10--40 Mm.}
\label{f:compare}
\end{figure*}

We have designed a ``flaring polarity inversion line'' (FPIL) mask to demarcate the AR core field, where most free energy resides (Figures~\ref{f:compare}(a)-(c)). We first identify the polarity inversion line (PIL) pixels from a smoothed vertical field $B_z$ map, and dilate them with a circular kernel (radius $r=3.5$ Mm). Then, we isolate flare ribbons from the 1600 {\AA} image (above 700 DN s$^{-1}$) taken near the flare peak by the Atmospheric Imaging Assembly \citep[AIA;][]{lemen2012} aboard \textit{SDO} and dilate them with a large kernel. The intersection of the dilated PIL and flare ribbons constitutes our FPIL mask. It resembles the mask in \citet{schrijver2007}, but includes only the part directly involved in a particular flare. Our conclusions are not affected if we adjust the mask width ($2r$) between 5 and 15 Mm.

%%%%%%%%%%%%%%%

%%%%%%%%%%%%%%%

In the largest sunspot umbrae, the HMI inversion module sometimes returns unreasonable field values with high formal errors. For example, a small patch of abnormally weak $B_z$ appeared at the center of the negative sunspot in AR 12192 (Figure~\ref{f:compare}(a)). The reason for these ``bad pixels'' is not fully understood; it appears to be the combined effect of low intensity, extremes in the \textit{SDO} orbital velocity, and limitations of the inversion technique. To estimate the adverse effect on our analysis, we identify these pixels by setting empirical thresholds on the formal errors and smoothly interpolate over them using the data nearby. The difference between the original and the interpolated data is 4$\%$ for $\Phi$, 1$\%$ for modeled $E_n$ and $E_p$, and 8$\%$ for $E_f$ (median in time). None affects our conclusions.

%%%%%%%%%%%%%%%%%%%

\section{Results}
\label{sec:results}

\subsection{Pre-Flare Conditions}
\label{subsec:preflare}

We summarize the pre-flare magnetic conditions and the flare-related changes for the three ARs in Table~\ref{t:keys}. Various indices shown to be useful indicators for flare and CME activity are computed and can be classified as either extensive or intensive, following \citet{welsch2009}. Extensive indices generally scale with AR size (e.g., totals) while intensive indices do not (e.g., means). We have the following observations for AR 12192 regarding the pre-flare conditions, in comparison with ARs 11429 and 11158.

\begin{enumerate}[parsep=0ex,partopsep=0ex,itemsep=1ex,leftmargin=3mm]

\item Its global, extensive-type indices are significantly greater. These include sunspot area, total magnetic flux $\Phi$, electric current $I$, magnetic energy $E_p$ and $E_n$, and FPIL mask size.

\item Its extensive-type indices in the core field are comparable to the other two ARs. These include the $R$ parameter \citep[a free energy proxy,][]{schrijver2007}, $\Phi$ and $I$ within the FPIL mask, and free energy $E_f$.

\item Its intensive-type indices, particularly those regarding AR non-potentiality in the core field, are significantly weaker. These include rms field $B$, rms electric current density $J$, mean shear angle, mean torsional parameter $\alpha$ within the FPIL mask, and relative free energy $E_f/E_p$. The net current helicity of the extensive type is small too. This is nicely illustrated by the vertically integrated $J$ maps from the NLFFF model (Figures~\ref{f:compare}(d)-(f)).
 
\item Its background field straddling the FPIL is significantly stronger. In the PF model, $B_h$ of AR 12192 is stronger in low corona ($z\approx42$ Mm); the relative strength with respect to the near-surface ($z\approx2$ Mm) value is higher too. Below 120 Mm, $B_h$ decreases much slower with height, leading to a lower decay index $n$ (Figures~\ref{f:compare}(g)-(i)); $n$ does not reach 1.5 until a large altitude, so the torus instability is less likely to set in. We obtain the same conclusions using the NLFFF model.

\end{enumerate}

%%%%%%%%%%%%%%%

\begin{figure*}[t!]
\centerline{\includegraphics{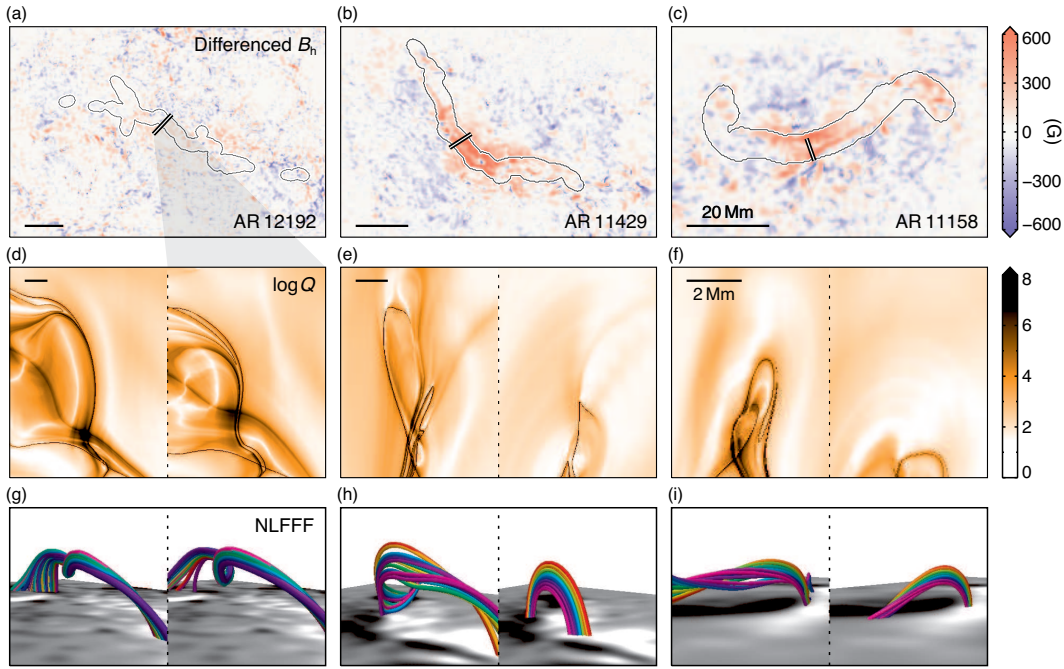}}
\caption{Magnetic changes over 1 hr. (a)-(c) Differenced photospheric $B_h$. Contours show the FPIL masks. (d)-(f) Logarithm squashing factor $Q$ on a vertical cut in the NLFFF model, before (left) and after (right) each event. The location of the cut is marked by a short double line in (a)-(c). (g)-(i) Selective field lines demonstrating the connectivity changes before (left) and after (right). The horizontal and vertical axes in (a)-(f) have the same scale; height in (g)-(i) is stretched for clarity.}
\label{f:change}
\end{figure*}

%%%%%%%%%%%%%%%

\subsection{Flare-Related Changes}
\label{subsec:change}

We evaluate the flare-related changes using selected indices from the last frame before the flare onset and the frame 1 hr after. The actual change near the FPIL is permanent and likely occurs on a time scale of minutes \citep{sudol2005}. AR 12192 is distinctive in the following aspects.

\begin{enumerate}[parsep=0ex,partopsep=0ex,itemsep=1ex,leftmargin=3mm]

\item Its photospheric field change is significantly weaker. In the two eruptive cases, $B_h$ increased by hundreds of gauss in the AR core; such change is not present in AR 12192 (Figures~\ref{f:change}(a)-(c)). To assess the significance, we take the difference between pairs of $B_h$ maps and compute the rms ($\sigma$) in the FPIL. The pairs are taken both before or after the flare, thus represent a baseline from secular evolution. We find that the changes of mean $B_h$ in the FPIL of ARs 11429 and 11158 reach 2.9$\sigma$ and 2.2$\sigma$ respectively (over 5$\sigma$ in the central part), while for AR 12192 the change is 0.2$\sigma$. Similarly, the change of $F_z \propto \sum(B_h^2-B_z^2)$, which possibly correlates with the ejecta momentum in eruptive flares \citep{fisher2012}, is much smaller in AR 12192.

\item Its inferred coronal field change is smaller. We compute the squashing factor $Q$ from the NLFFF model on a vertical cross section (Figures~\ref{f:change}(d)-(f)). In all three ARs, the pre-flare high-$Q$ patterns indicate the existence of twisted magnetic structure (Figures~\ref{f:change}(g)-(i)). In the two eruptive cases, such structure largely disappears  after eruption, indicating fundamental topological change. The change is less significant for AR 12192.

\end{enumerate}

We note that the changes of other indices in Table~\ref{t:keys} may exhibit a variety of behaviors\footnote{See \url{http://purl.stanford.edu/ss540kq5576} for an extended table.}. For example, $\Phi$ changes very little, while current helicity increases significantly in all three ARs. We defer investigation of these behaviors to future work.

%%%%%%%%%%%%%%%%%%%

\section{Interpretation and Discussion}
\label{sec:discussion}

AR 12192 exhibits weaker non-potentiality and stronger overlying field in the core region, consistent with previously studied confined cases. However, the region's flare productivity is extraordinary; most of its extensive properties are comparable to or greater than the eruptive ARs 11429 and 11158. The estimated free energy is enough to power multiple X-class flares. We thus argue against global, absolute measures as the \textit{physical} controlling factor of AR eruptiveness, and suggest using instead some \textit{relative} measure that quantifies the ratio of magnetic non-potentiality to the restriction of background field. The possibly different controlling factors of flare and CME productivity provide a viable explanation to the strange behavior of AR 12192.

The exact formula of such a relative measure is unknown but may be explored through surveys. The prediction capability, of course, remains probabilistic. If the measure is relatively low, any flare from the AR is likely confined. If it is high, major flares are likely eruptive; smaller flares, however, can still be confined.

HMI observations have confirmed sudden and permanent photospheric field change as a common feature in large flares \citep{sun2012,wangs2012}. $B_h$ generally strengthens along the FPIL, consistent with the magnetic ``implosion'' scenario where coronal loops contract in response to the reduction of magnetic pressure from energy release \citep{hudson2000}.

Along this line, \citet{fisher2012} argue that one can use the photospheric integral of Maxwell stress tensor to estimate the total Lorentz force in a carefully selected volume. The temporal integral of its change, which results from the change of the photospheric field, can represent the Lorentz force impulse that provides the CME momentum. As the ejecta momentum is small in the confined X3.1 flare, the weak change of $B_h$ seems to support the argument.

It is unclear whether all confined flares exhibit similar small changes, although such is consistent with theoretical expectations. In eruptive flares, the CME bodily removes magnetic helicity by ejecting twisted flux ropes into interplanetary space, resulting in less-sheared post-flare loops \citep{priest2002}. In confined events, however, helicity is largely conserved. The post-flare topological complexity is expected to differ less, which is what we find.

The EUV and X-ray observations of AR 12192 show some unorthodox features. For example, they have relatively long X-ray duration: the X3.1 confined flare lasted 66 minutes, much longer than the two eruptive flares in ARs 11429 and 11158 (Table~\ref{t:keys}); the median duration of all confined flares above M3 is 53 minutes. An estimate of the non-thermal electron energy for an earlier, confined X1 flare yields $\sim$$10^{32}$ erg, significantly larger than that of a typical, eruptive X1 flare \citep{thalmann2015}. In addition, the large overlying loops that connect to the southwest plage region appear to be continuously energized \citep[Figure~\ref{f:context}(c); see also,][]{liur2014}. These features suggest a different energy partition in confined events. A detailed analysis, similar to those performed on eruptive flares \cite[e.g.,][]{emslie2004,feng2013}, will be worthwhile and may shed light on the triggering mechanism.

We note that our proposed relative measure is different from the existing, \textit{statistical} CME predictors that are extensive in nature \citep[e.g., proxies for $E_f$,][]{falconer2006}. There is no direct conflict, because AR 12192 is clearly a statistical outlier. In fact, any new index must statistically correlate with established CME predictors to have any predictive power. Since the $E_f$ proxies also correlate with $\Phi$ \citep{falconer2002}, our relative measure should not be scale-free.

The HMI database has accumulated 4.5 years of vector field data, totaling over 4000 strong-field regions and 1.7 million records. A comprehensive statistical study is now possible and has been performed for flare likelihood \citep{bobra2015}. A similar survey can help clarify whether AR 12192 is a representative case for confined flares and can provide useful insights to CME forecasting.

%%%%%%%%%%%%%%%

\acknowledgments
We are grateful to Thomas Wiegelmann for the extrapolation code, Seiji Yashiro for the CME-flare statistics, and Anna Malanushenko for the loop-fitting software. This work is supported by NASA contract NAS5-02139, awards NNX11AJ65G, NNX13AK39G, NSF awards AGS-1321474, and AGS-1249150. The \textit{SDO} data are courtesy of NASA, the \textit{SDO}/HMI and AIA science teams. Magnetic field lines are visualized using VAPOR.

{\it Facilities:} \facility{\textit{SDO}}.\\

%%%%%%%%%%%%%%%%%%%

\end{CJK}

\bibliographystyle{yahapj}
\bibliography{cmeless}

%%%%%%%%%%%%%%%

%\clearpage
%[width=0.6\textwidth]

%%%%%%%%%%%%%%%%%%%

\end{document}